\documentclass[journal]{IEEEtran}
\usepackage{cite}
\usepackage{amsmath,amssymb,amsfonts}
\usepackage{graphicx}
\usepackage{textcomp}
\usepackage{xcolor}
\usepackage{booktabs}
\usepackage{multirow}
\usepackage{hyperref}
\usepackage{url}
\def\BibTeX{{\rm B\kern-.05em{\sc i\kern-.025em b}\kern-.08em
    T\kern-.1667em\lower.7ex\hbox{E}\kern-.125emX}}

\newcommand{\figmiss}[1]{}
\newcommand{\expmiss}[1]{}
\begin{document}

\title{WiSER: A Wireless Scene Encoder for Geometry-Grounded Multi-View Wireless Prediction}

\author{
Jing Qiao,
Yiyang Guo,
and Hao Ye%
\thanks{
Jing Qiao, Yiyang Guo, and Hao Ye are with the Department of Electrical and Computer Engineering,
University of California Santa Cruz, Santa Cruz, CA 95064 USA
(e-mail: \{jqiao5, yguo173, yehao\}@ucsc.edu).
}%
}

\markboth{IEEE Journal Submission}%
{Qiao \MakeLowercase{\textit{et al.}}: WiSER: A Wireless Scene Encoder for Geometry-Grounded Multi-View Wireless Prediction}

\maketitle

\begin{abstract}
Indoor wireless propagation is governed by the interaction among
three-dimensional (3D) scene geometry, radio-material properties, and
transmitter and receiver configuration, which jointly determine both aggregate coverage behavior and path-level multipath structure. However, most learning-based site-specific prediction methods are designed for a single wireless representation, such as radiomap estimation or channel impulse response (CIR) prediction, and therefore do not explicitly exploit the propagation structure shared across heterogeneous wireless views. This paper introduces WiSER, a Wireless Scene Encoder for joint radiomap and multipath CIR prediction. WiSER maps a sparse voxel representation of an indoor scene and a transmitter location into a transmitter-conditioned sparse 3D scene memory, which is queried by two structure-aware decoders: a ray-corridor decoder for dense receiver-plane path-gain prediction and a Detection Transformer (DETR)-style set decoder for variable-cardinality delay and power tap prediction. To train and evaluate this setting, we construct a co-registered indoor scene and wireless dataset pipeline using ScanNet++ indoor scenes and Sionna Ray Tracing, producing aligned sparse voxel inputs, dense radiomap labels, and unordered multipath CIR tap sets under a common
coordinate frame and propagation configuration. Experimental results show that WiSER outperforms scene-specific radiomap baselines and substantially improves matched delay and power prediction over reference CIR baselines. These results suggest that transmitter-conditioned sparse 3D scene representations can serve as reusable wireless scene encoders for heterogeneous propagation queries, providing a geometry-grounded step toward representation learning and foundation-model development for AI-native wireless systems.
\end{abstract}


\section{Introduction}
\label{sec:intro}

\IEEEPARstart{I}{ndoor} wireless propagation is a scene-conditioned physical
process governed by three-dimensional (3D) geometry, radio-material properties,
and transmitter and receiver configuration. These factors jointly determine both
aggregate coverage behavior and link-level multipath structure. Classical
stochastic channel models provide compact abstractions of path loss, fading,
delay spread, and angular statistics, but they largely abstract away
site-specific environmental structure~\cite{cost2100,quadriga,nyusim}. In
contrast, deterministic ray-tracing and digital-twin simulators explicitly
model geometry, materials, blockage, reflection, diffraction, scattering, and
transmitter and receiver placement, and can therefore generate physically
interpretable site-specific channels~\cite{sionna,sionna_rt}. However,
repeated ray tracing remains costly when predictions are required across many
scenes, transmitter locations, receiver locations, carrier configurations, and
wireless query types.

Learning-based wireless prediction offers a data-driven surrogate for such
site-specific simulation. Prior work has studied radiomap and coverage-map
prediction, channel-knowledge-map construction, and spectrum cartography for
aggregate spatial prediction
\cite{radiomap_dl_1,radio_map_estimation,ckm_i2i,ckm_tutorial}, as well as
channel state information (CSI), channel feedback, channel impulse response
(CIR), tapped-delay-line parameters, and other path-level quantities for
link-specific prediction~\cite{csi_dl_1,csinet_plus,deepcmc}. These studies
show that neural models can approximate useful wireless quantities from data.
However, most existing methods are organized around a single site, output
representation, or wireless task. A radiomap predictor is typically designed
for dense received-power or path-gain estimation, whereas a CIR predictor is
designed for link-level multipath prediction. This separation leaves open the
question of whether heterogeneous wireless views can be decoded from a shared
representation of the same physical scene.

The emerging interest in wireless foundation models further motivates this
question. A central goal of this direction is to learn reusable representations
that can support multiple downstream wireless tasks rather than training an
isolated model for each query type. Much of the recent progress has focused on
signal-domain representations, such as channel tensors, channel sequences, or
space, time, and frequency measurements~\cite{wireless_fm_perspective,lwm2024,wifo2024}.
For site-specific wireless systems, however, the physical environment is also a
fundamental conditioning signal. This paper studies a complementary
geometry-grounded direction: whether a sparse 3D representation of the radio
scene can serve as a reusable wireless scene encoder for heterogeneous
propagation queries.

We instantiate this question with two complementary wireless prediction tasks:
radiomap estimation over receiver planes and multipath CIR prediction for
transmitter and receiver links. The two views have distinct output structures.
Radiomap prediction is dense spatial field regression, whereas multipath CIR
prediction requires unordered delay and power tap sets with variable
cardinality. A shared model must therefore preserve propagation-relevant 3D
structures, such as blockages, openings, and reflective surfaces; condition the
scene representation on transmitter placement; and decode each wireless view
with an output structure matched to its physical representation. These
requirements also call for co-registered supervision that aligns scene
geometry, transmitter and receiver coordinates, radiomap labels, and path-level
CIR labels under a common coordinate system and propagation model.

To address these requirements, we construct a co-registered indoor scene and
wireless dataset pipeline and propose WiSER, a Wireless Scene Encoder that
learns a reusable transmitter-conditioned scene representation for joint
radiomap and multipath CIR prediction. Starting from ScanNet++ indoor
scenes~\cite{scannetpp}, each environment is converted into a sparse 3D voxel
representation for learning and a Sionna Ray Tracing scene for label
generation~\cite{sionna_rt}. In contrast to wireless datasets that primarily
target fixed channel matrices, sensing-assisted links, or stand-alone radiomap
outputs~\cite{deepmimo,deepsense6g,m3sc,radiodiff3d}, our pipeline
co-registers two heterogeneous wireless views for the same 3D scene. Dense
radiomap labels and multipath CIR tap sets are generated in the same coordinate
frame, providing aligned coverage-level and path-level supervision for the same
physical scene.

WiSER maps a voxelized indoor scene and transmitter location into a shared
transmitter-conditioned sparse 3D scene representation. Two task-specific
decoders then query this representation. The radiomap decoder uses
ray-corridor scene access to predict dense receiver-plane path-gain fields. The
CIR decoder uses a Detection Transformer (DETR)-style set prediction head with
Hungarian matching to predict variable-cardinality delay and power tap
sets~\cite{detr}. At the operator level, WiSER uses sparse transformer
primitives released with TRELLIS-2~\cite{trellis2}; the
transmitter-conditioning interface, scene-memory design, and wireless
prediction decoders are developed in this work.

We evaluate WiSER across indoor scenes using radiomap and multipath CIR
prediction tasks. The results show that a shared sparse 3D scene representation
can support both aggregate coverage-level prediction and path-level channel
prediction within a single architecture. These results position WiSER as a
geometry-grounded step toward reusable wireless representations: instead of
learning a separate predictor for each wireless output, the model learns a
transmitter-conditioned scene memory that can be queried by multiple
propagation views under controlled ray-tracing supervision.

The main contributions of this paper are summarized as follows:
\begin{enumerate}
\item We formulate radiomap prediction and multipath CIR prediction as two
complementary views of the same scene-conditioned propagation process,
providing a concrete setting for studying reusable geometry-grounded
representations for wireless prediction.

\item We construct a ScanNet++/Sionna-based co-registered scene--wireless
dataset pipeline that aligns sparse 3D scene geometry, transmitter and receiver
configurations, dense radiomap labels, and path-level multipath CIR labels in a
common coordinate frame.

\item We introduce WiSER, a transmitter-conditioned sparse 3D scene encoder
with two structure-aware wireless decoders: a ray-corridor radiomap decoder for
dense path-gain prediction and a DETR-style CIR set decoder for unordered
delay--power tap prediction.

\item We evaluate WiSER against radiomap and CIR baselines and conduct ablation
studies on the main architectural and training components, showing that a
single transmitter-conditioned sparse 3D scene representation can support both
dense coverage-level and sparse path-level wireless prediction under controlled
ray-tracing supervision.
\end{enumerate}

The remainder of this paper is organized as follows.
Section~\ref{sec:related} reviews related work.
Section~\ref{sec:background} formulates the geometry-conditioned multi-view
wireless prediction problem.
Section~\ref{sec:methods} presents the proposed architecture and training
objective.
Section~\ref{sec:dataset_generation} describes the dataset-generation pipeline.
Section~\ref{sec:results} reports experimental results and ablation studies.
Section~\ref{sec:conclusion} concludes the paper.
\section{Related Work}
\label{sec:related}

This work lies at the intersection of learning-based wireless prediction,
wireless representation learning, and geometry-grounded channel modeling. We
review these areas with emphasis on a common limitation of prior work: most
methods are designed for a single wireless output representation rather than a
shared scene-conditioned representation supporting heterogeneous wireless views.

\textbf{Task-specific wireless prediction and output structure.}
Learning-based wireless prediction has been widely studied as a surrogate for
measurement-intensive or simulation-intensive site-specific channel modeling.
One line of work predicts aggregate spatial quantities, including radiomaps,
coverage maps, channel-knowledge maps, spectrum-cartography representations,
and 3D radio maps~\cite{radiomap_dl_1,radio_map_estimation,ckm_i2i,
ckm_tutorial,radiodiff3d,radiogen3d}. These methods estimate received power,
path gain, or related channel-quality metrics over spatial grids and are
naturally formulated as dense regression problems. Another line of work focuses
on link-level channel quantities, such as CSI,
channel feedback, CIR, and tapped-delay-line
parameters~\cite{csi_dl_1,csinet_plus,deepcmc,sitespecific_channel}. These
outputs have a different structure from radiomaps: multipath CIRs are sparse
unordered sets whose cardinality varies with the transmitter--receiver link.
Fixed-vector regression is therefore not well matched to path-level multipath
prediction, which motivates permutation-invariant supervision such as
DETR-style set prediction and later query-based
detection variants~\cite{detr,deformable_detr}. Existing
wireless studies typically optimize separate models for aggregate coverage
prediction and path-level channel prediction. In contrast, this paper treats
radiomap and multipath CIR prediction as two supervised views of the same
scene-conditioned propagation process.

\textbf{Wireless representation learning, datasets, and multi-task learning.}
Recent work on wireless representation learning and wireless foundation models
aims to move beyond task-specific predictors toward reusable models that
support multiple downstream wireless tasks~\cite{bommasani2021foundation,wireless_fm_perspective}. In
parallel, datasets such as DeepMIMO, DeepSense 6G, and M3SC have made
ray-tracing or measurement-based wireless learning more reproducible for
channel matrices, sensing-assisted links, and multi-modal communication
benchmarks~\cite{deepmimo,deepsense6g,m3sc}. Most existing efforts emphasize
signal-domain observations, such as CSI tensors, channel sequences, or
space--time--frequency measurements. For site-specific wireless prediction,
however, the physical scene provides an additional source of shared structure
across tasks. This motivates a complementary geometry-grounded direction, in
which an environment representation is learned once and queried for multiple
wireless outputs. The joint setting is also related to multi-task learning,
where shared representations, heterogeneous losses, and task interference have
been studied extensively~\cite{caruana1997,kendall2018}. Our focus is therefore complementary to signal-centric wireless
representation learning: we study whether the physical scene itself can provide
a reusable representation for heterogeneous wireless queries.

\textbf{Geometry-grounded wireless modeling.}
The physical environment is central to site-specific wireless prediction.
Classical channel models and simulators provide compact or deterministic
descriptions of propagation but differ in how explicitly they represent
scene geometry~\cite{cost2100,quadriga,nyusim,sionna,sionna_rt}. Deterministic
ray-tracing simulators, such as Sionna Ray Tracing, provide physically
interpretable supervision by modeling interactions among radio waves, scene
geometry, materials, and transmitter--receiver placement~\cite{sionna_rt}.
Such simulators can generate path gains, radiomaps, delays, angles, and
multipath coefficients, but repeated simulation can be computationally costly.
This has motivated neural surrogates that condition on environmental
information, including two-dimensional floor plans, occupancy maps, satellite
or map imagery, point clouds, neural fields, Gaussian-splatting
representations, and 3D scene inputs~\cite{nerf,3dgs,
nerf2,newrf,rf3dgs,winert}. Sparse 3D representations are
attractive for indoor scenes because most of the volume is empty, while
propagation-relevant structures are concentrated on occupied surfaces and
objects. Sparse convolutional networks, Minkowski convolutions, and point or
voxel transformers process only occupied elements and have become common
building blocks for scalable 3D learning~\cite{minkowski,point_transformer_v2,trellis2}. WiSER uses sparse
transformer primitives from TRELLIS-2 as operator-level building blocks, but
the transmitter-conditioned scene interface and wireless decoders are specific
to the propagation prediction problem. Unlike prior geometry-aware surrogates
optimized for a single wireless output, our dataset and model are explicitly
co-registered across dense radiomap labels and path-level CIR labels.

\section{Problem Formulation}
\label{sec:background}

We formulate geometry-grounded wireless prediction as a scene-conditioned
channel learning problem. Each indoor environment is represented by a sparse
3D description of its physical structure, and wireless
quantities are queried by specifying transmitter and receiver configurations.
The objective is to learn a shared scene-conditioned representation that
supports two complementary views of the same propagation process: a dense
radiomap and a path-level multipath CIR.

\subsection{Scene-Conditioned Channel Model}

Let $\mathcal{S}$ denote an indoor scene characterized by geometry, semantic
structure, and radio-material properties. For a transmitter located at
$\mathbf{x}_{\rm t}\in\mathbb{R}^{3}$ and a receiver located at
$\mathbf{x}_{\rm r}\in\mathbb{R}^{3}$, the site-specific baseband channel is
modeled as
\begin{equation}
    h_{\mathcal{S}}(\tau;\mathbf{x}_{\rm t},\mathbf{x}_{\rm r})
    =
    \sum_{\ell\in
    \mathcal{L}_{\mathcal{S}}(\mathbf{x}_{\rm t},\mathbf{x}_{\rm r})}
    \alpha_{\ell}\delta(\tau-\tau_{\ell}),
    \label{eq:channel_model}
\end{equation}
where $\mathcal{L}_{\mathcal{S}}(\mathbf{x}_{\rm t},\mathbf{x}_{\rm r})$ is
the index set of valid propagation paths between $\mathbf{x}_{\rm t}$ and
$\mathbf{x}_{\rm r}$, $\alpha_{\ell}\in\mathbb{C}$ is the complex coefficient of path $\ell$,
$\tau_{\ell}$ is its propagation delay, and $\delta(\cdot)$ denotes the Dirac
delta function. Under a fixed carrier frequency, antenna configuration, and
propagation model, $\alpha_{\ell}$ captures the effects of antenna response,
material interaction, path loss, reflection, diffraction, scattering, and other
ray-tracing assumptions.

The physical scene is represented through a sparse voxelized description of the
radio environment. This representation is not tied to a specific sensing
modality; it may be constructed from meshes, point clouds, RGB-D scans, LiDAR maps, semantic
occupancy maps, or other 3D scene sources. Let $c_v>0$ denote
the voxel edge length in meters. Given a world coordinate frame with origin
$\mathbf{x}_{0}\in\mathbb{R}^{3}$, the 3D space is discretized into cubic
voxels of side length $c_v$, and only voxels associated with occupied surfaces
or objects are retained. The resulting sparse voxel set is
\begin{equation}
    \mathcal{V}_{\mathcal{S}}
    =
    \left\{
    (\mathbf{p}_{i},\mathbf{f}_{i})
    \right\}_{i=1}^{N_{\mathcal{S}}},
    \qquad
    \mathbf{p}_{i}\in\mathbb{Z}^{3},\quad
    \mathbf{f}_{i}\in\mathbb{R}^{C_v},
    \label{eq:sparse_voxel}
\end{equation}
where $\mathbf{p}_{i}=[p_{i,x},p_{i,y},p_{i,z}]^{\mathsf{T}}$ is the integer
grid coordinate of the $i$-th occupied voxel, $\mathbf{f}_{i}$ is its feature
vector, $C_v$ is the voxel-feature dimension, and $N_{\mathcal{S}}$ is the
number of occupied voxels in scene $\mathcal{S}$. The physical center of voxel
$i$ is
\begin{equation}
    \mathbf{x}_{i}
    =
    \mathbf{x}_{0}
    +
    c_v
    \begin{bmatrix}
    p_{i,x}+\frac{1}{2} \\
    p_{i,y}+\frac{1}{2} \\
    p_{i,z}+\frac{1}{2}
    \end{bmatrix}.
    \label{eq:voxel_center}
\end{equation}
The integer coordinate $\mathbf{p}_{i}$ specifies the sparse-grid index used by
the 3D encoder, whereas $\mathbf{x}_{i}$ specifies the physical location of the
voxel center in meters. The feature vector $\mathbf{f}_{i}$ may encode
occupancy, color or intensity, semantic category, surface or object type, and
material-related attributes, depending on the available scene source. The sparse
set $\mathcal{V}_{\mathcal{S}}$ therefore preserves the 3D locations and
attributes of potential blockers, reflectors, and scatterers while avoiding
dense computation over empty space. All voxel centers, transmitter locations,
receiver locations, radiomap cells, and CIR labels are expressed in the same
world coordinate system.

\subsection{Radiomap View}

The radiomap provides an aggregate spatial view of the channel over a receiver
plane. For a plane at height $z$, define the receiver lattice as
\begin{equation}
    \mathcal{R}_{z}
    =
    \left\{
    \mathbf{x}_{u,v}(z)
    \;\middle|\;
    1\leq u\leq N_y(z),\;
    1\leq v\leq N_x(z)
    \right\},
    \label{eq:rx_lattice}
\end{equation}
where $\mathbf{x}_{u,v}(z)\in\mathbb{R}^{3}$ is the receiver coordinate
associated with grid cell $(u,v)$, and $N_y(z)$ and $N_x(z)$ denote the row and
column counts of the receiver lattice. For transmitter $\mathbf{x}_{\rm t}$, the
radiomap label is defined as the path-gain field
\begin{equation}
    G_{\mathcal{S}}(\mathbf{x}_{\rm t},z)[u,v]
    =
    10\log_{10}
    \left(
    \sum_{\ell\in
    \mathcal{L}_{\mathcal{S}}(\mathbf{x}_{\rm t},\mathbf{x}_{u,v}(z))}
    |\alpha_{\ell}|^{2}
    \right),
    \label{eq:radiomap_def}
\end{equation}
for valid receiver cells. This quantity is a path gain in dB under the
normalization convention used for the path coefficients. Cells without valid
ray-tracing labels are indicated by a binary mask
$M_{\mathcal{S}}(\mathbf{x}_{\rm t},z)\in\{0,1\}^{N_y(z)\times N_x(z)}$ and are
excluded from the training loss and evaluation metrics.

The radiomap prediction task is written as
\begin{equation}
    \widehat{G}_{\theta}(\mathbf{x}_{\rm t},z)
    =
    f_{\theta}^{\rm rm}
    \left(
    \mathcal{V}_{\mathcal{S}},\mathbf{x}_{\rm t},z
    \right),
    \label{eq:rm_prediction}
\end{equation}
where
$\widehat{G}_{\theta}(\mathbf{x}_{\rm t},z)\in\mathbb{R}^{N_y(z)\times N_x(z)}$
is the predicted dense path-gain map.

\subsection{Multipath CIR View}

The multipath CIR provides a path-level view of the same propagation process.
For a transmitter--receiver pair
$(\mathbf{x}_{\rm t},\mathbf{x}_{\rm r})$, the supervised CIR target is
represented as an unordered set of delay--power taps,
\begin{equation}
    \mathcal{Y}_{\mathcal{S}}(\mathbf{x}_{\rm t},\mathbf{x}_{\rm r})
    =
    \left\{
    \mathbf{y}_{m}
    \right\}_{m=1}^{M},
    \qquad
    \mathbf{y}_{m}=(\tau_m,\gamma_m),
    \label{eq:cir_set}
\end{equation}
where $M$ is the number of valid taps, $\tau_m$ is the delay of tap $m$, and
\begin{equation}
    \gamma_m = 10\log_{10}|\alpha_m|^2
    \label{eq:path_power}
\end{equation}
is the tap power in decibels. Since propagation paths do not have a canonical
ordering, $\mathcal{Y}_{\mathcal{S}}$ is treated as a finite set rather than a
sequence. CIR prediction is therefore a permutation-invariant
variable-cardinality prediction problem.

The CIR prediction task is
\begin{equation}
    \widehat{\mathcal{Y}}_{\theta}(\mathbf{x}_{\rm t},\mathbf{x}_{\rm r})
    =
    f_{\theta}^{\rm cir}
    \left(
    \mathcal{V}_{\mathcal{S}},
    \mathbf{x}_{\rm t},
    \mathbf{x}_{\rm r}
    \right),
    \label{eq:cir_prediction}
\end{equation}
where the output is an unordered set of predicted delay--power taps.

\subsection{Multi-View Learning Objective}

The training data consist of co-registered radiomap and CIR samples generated
from the same scene geometry and propagation configuration. Let
$\mathcal{D}_{\rm rm}$ and $\mathcal{D}_{\rm cir}$ denote the radiomap and CIR
datasets, respectively:
\begin{equation}
    \mathcal{D}_{\rm rm}
    =
    \left\{
    \left(
    \mathcal{V}_{\mathcal{S}},
    \mathbf{x}_{\rm t},
    z,
    G_{\mathcal{S}}(\mathbf{x}_{\rm t},z),
    M_{\mathcal{S}}(\mathbf{x}_{\rm t},z)
    \right)
    \right\},
    \label{eq:rm_dataset}
\end{equation}
\begin{equation}
    \mathcal{D}_{\rm cir}
    =
    \left\{
    \left(
    \mathcal{V}_{\mathcal{S}},
    \mathbf{x}_{\rm t},
    \mathbf{x}_{\rm r},
    \mathcal{Y}_{\mathcal{S}}(\mathbf{x}_{\rm t},\mathbf{x}_{\rm r})
    \right)
    \right\}.
    \label{eq:cir_dataset}
\end{equation}
Co-registration requires the scene representation, transmitter and receiver
coordinates, radiomap labels, and CIR labels to be defined in the same
coordinate frame and generated under the same propagation setting.

The multi-view learning objective is
\begin{align}
    \min_{\theta}\; \mathcal{J}(\theta)
    &=
    \mathbb{E}_{\mathcal{D}_{\rm rm}}
    \left[
    \mathcal{L}_{\rm rm}
    \left(
    \widehat{G}_{\theta},
    G_{\mathcal{S}},
    M_{\mathcal{S}}
    \right)
    \right]
    \nonumber\\
    &\quad+
    \lambda_{\rm cir}
    \mathbb{E}_{\mathcal{D}_{\rm cir}}
    \left[
    \mathcal{L}_{\rm cir}
    \left(
    \widehat{\mathcal{Y}}_{\theta},
    \mathcal{Y}_{\mathcal{S}}
    \right)
    \right],
    \label{eq:joint_objective}
\end{align}
where $\mathcal{L}_{\rm rm}$ is a masked dense regression loss for radiomap
prediction, $\mathcal{L}_{\rm cir}$ is a permutation-invariant set loss for
multipath CIR prediction, and $\lambda_{\rm cir}$ controls the relative weight
of the CIR supervision. This formulation separates the shared
scene-conditioned representation problem from the view-specific decoder and
loss designs introduced in Section~\ref{sec:methods}.

\section{Proposed Method}
\label{sec:methods}

WiSER is built around a transmitter-conditioned sparse scene memory, which
serves as a reusable interface between the 3D radio environment and multiple
wireless prediction views. Given a sparse voxel representation of a scene and a
transmitter location, the shared scene encoder constructs a memory that
combines local geometric evidence with transmitter-dependent propagation
context. This memory is then queried by two structurally different wireless
decoders: a ray-corridor radiomap decoder for dense receiver-plane path-gain
prediction and a multipath CIR set decoder for
unordered delay--power tap prediction.

The key distinction from a conventional multi-task architecture is that the
shared representation is not merely a generic backbone feature. It is explicitly
conditioned on transmitter placement and is designed to be queried by decoders
whose memory access patterns and losses match the physical structure of their
outputs. Dense radiomap prediction requires spatially coherent field decoding,
whereas multipath CIR prediction requires permutation-invariant
variable-cardinality set decoding. WiSER preserves this
distinction while amortizing the scene-and-transmitter encoding across both
wireless views.

\begin{figure*}[t]
  \centering
  \includegraphics[width=\textwidth]{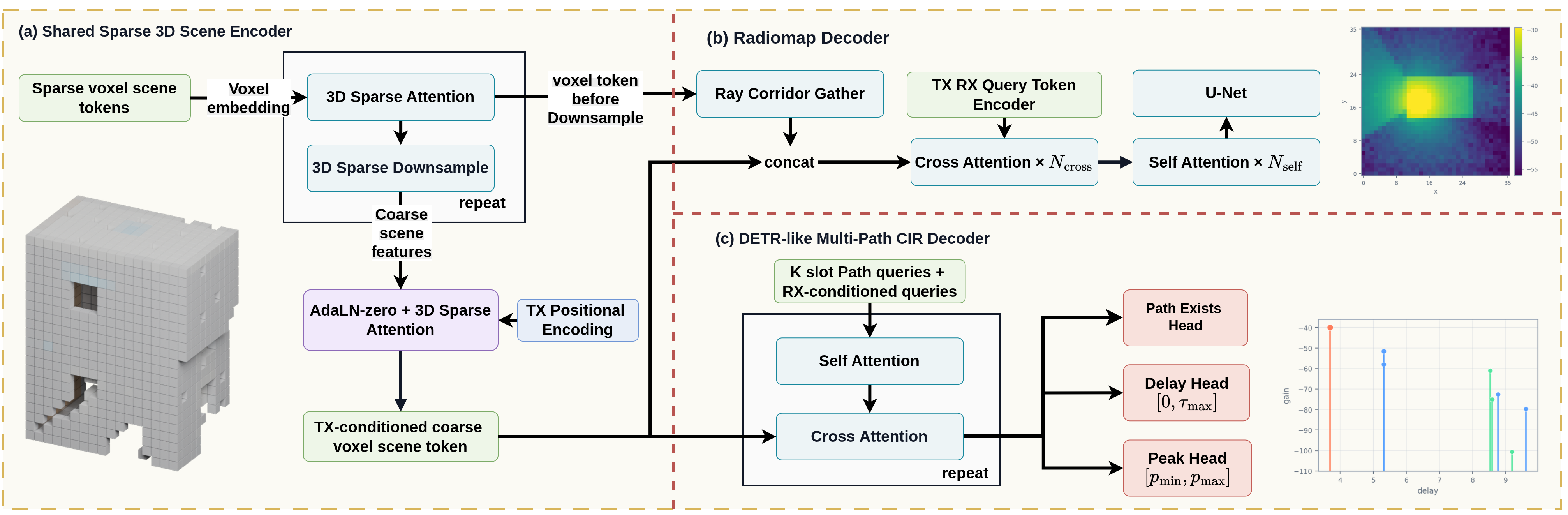}
  \caption{Overall WiSER architecture. (a) The sparse voxel scene tokens are
  encoded by repeated sparse attention and sparse downsampling, and the coarsest
  scene tokens are modulated by transmitter positional encoding to form a
  transmitter-conditioned sparse scene memory. (b) The radiomap decoder gathers
  ray-corridor scene tokens, combines them with transmitter--receiver query
  tokens, and predicts a dense receiver-plane path-gain map. (c) The CIR decoder
  uses receiver-conditioned path queries and DETR-like set decoding to predict
  existence, delay, and peak power for a variable-cardinality multipath tap set.}
  \label{fig:overview}
\end{figure*}

\subsection{Transmitter-Conditioned Sparse Scene Memory}
\label{sec:scene_memory}

Let $\mathcal{V}_{\mathcal{S}}$ denote the sparse voxel representation of
scene $\mathcal{S}$, and let $\mathbf{x}_{\rm t}$ denote the transmitter
location. The shared encoder maps the scene--transmitter pair to a
multi-resolution memory
\begin{equation}
    \mathbf{Z}_{\phi}
    =
    \mathbf{Z}_{\phi}
    \left(
    \mathcal{V}_{\mathcal{S}},
    \mathbf{x}_{\rm t}
    \right)
    =
    \left(
    \mathbf{X}_{\rm loc},
    \mathbf{M}_{\rm tx}
    \right),
    \label{eq:method_encoder_map}
\end{equation}
where $\mathbf{X}_{\rm loc}$ denotes the collection of fine and intermediate
sparse scene memories, and $\mathbf{M}_{\rm tx}$ denotes the
transmitter-conditioned global scene memory. The local memories retain
geometry near occupied surfaces and objects, while $\mathbf{M}_{\rm tx}$
summarizes the scene under the current transmitter placement.

The encoder operates only on occupied voxels. Sparse transformer blocks and
sparse downsampling aggregate local voxel features into a multi-resolution
scene representation, avoiding dense computation over empty indoor volume. This
is particularly suitable for propagation prediction because the relevant
structures are concentrated on surfaces and objects, such as walls, openings,
ceilings, furniture, reflectors, and blockages.

Transmitter conditioning is introduced at the coarsest level of the sparse
scene representation. Fine-scale memories are kept transmitter-agnostic so that
local geometry remains reusable across transmitter placements. In contrast, the
global memory is modulated by an embedding of $\mathbf{x}_{\rm t}$, obtained
from Fourier positional features followed by a multilayer perceptron (MLP). The
modulation is implemented through adaptive layer normalization zero
(AdaLN-zero) conditioning. This produces $\mathbf{M}_{\rm tx}$, a compact global memory that
encodes both the physical scene and the transmitter-dependent propagation
context. Sparse transformer and downsampling operators are implemented using
TRELLIS-2 primitives~\cite{trellis2}; the transmitter-conditioned scene-memory
interface and the wireless decoders are developed for WiSER. The encoder and
the two decoders are summarized together in Fig.~\ref{fig:overview}.

The radiomap and CIR decoders query the same memory:
\begin{align}
    \widehat{G}_{\theta}(\mathbf{x}_{\rm t},z)
    &=
    D^{\rm rm}_{\theta_{\rm rm}}
    \left(
    \mathbf{Z}_{\phi},z
    \right),
    \label{eq:rm_operator} \\
    \widehat{\mathcal{Y}}_{\theta}
    (\mathbf{x}_{\rm t},\mathbf{x}_{\rm r})
    &=
    D^{\rm cir}_{\theta_{\rm cir}}
    \left(
    \mathbf{Z}_{\phi},\mathbf{x}_{\rm r}
    \right).
    \label{eq:cir_operator}
\end{align}
The radiomap decoder is queried by a receiver-plane height $z$, while the CIR
decoder is queried by a receiver coordinate $\mathbf{x}_{\rm r}$.

\subsection{Ray-Corridor Radiomap Decoder}
\label{sec:radiomap_decoder}

Radiomap prediction is a dense receiver-plane decoding problem. For receiver
cell $(u,v)$ on plane $z$, let $\mathbf{x}_{u,v}(z)$ denote the receiver
coordinate. The transmitter--receiver geometry is represented as
\begin{equation}
    \mathbf{g}_{u,v}
    =
    \left[
    \mathbf{x}_{u,v}(z),
    \mathbf{x}_{\rm t},
    \boldsymbol{\Delta}_{u,v},
    d_{u,v}
    \right],
    \label{eq:txrx_geometry}
\end{equation}
where
$\boldsymbol{\Delta}_{u,v}
=
\mathbf{x}_{u,v}(z)-\mathbf{x}_{\rm t}$
and
$d_{u,v}
=
\|\boldsymbol{\Delta}_{u,v}\|_2$.
The vector $\mathbf{g}_{u,v}$ is embedded into a receiver query token
$\mathbf{q}_{u,v}$ using positional encodings and a multilayer perceptron.

\begin{figure}[t]
    \centering
    \includegraphics[width=0.5\columnwidth]{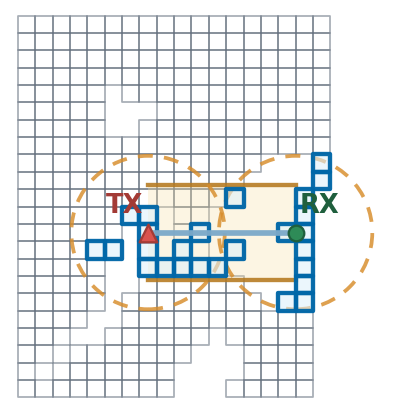}
    \caption{Ray-corridor feature gathering. (a) For a transmitter/receiver
    (TX/RX) query, the transmitter--receiver segment and endpoint neighborhoods define a compact
    candidate region in a sparse voxel slice. (b) A fixed-budget subset of
    nearby voxel cells is selected as the receiver-specific local memory. The
    schematic is shown in an $x$--$z$ slice for readability; the implemented
    selection is computed in the full 3D voxel coordinate frame.}
    \label{fig:ray_corridor}
\end{figure}

The central operation in the radiomap branch is ray-corridor gathering as shown in
Fig.~\ref{fig:ray_corridor}. Instead
of allowing every receiver query to attend to all occupied voxels, the decoder
selects a compact set of fine-scale scene tokens near the
transmitter--receiver segment and its endpoints. This operation acts as a
geometry-constrained attention sparsifier. It is not intended to reproduce
deterministic ray tracing; rather, it injects a weak propagation prior by
exposing each receiver query to likely blockers, openings, reflectors, and
nearby scattering structures.

Let $\mathbf{x}_{j}$ denote the physical center of fine-scale scene token $j$.
For receiver cell $(u,v)$, a corridor score
$s_{\rm corr}(\mathbf{x}_{j};\mathbf{x}_{\rm t},\mathbf{x}_{u,v})$ measures
the relevance of token $j$ to the transmitter--receiver corridor. The selected
local token indices are
\begin{equation}
    \mathcal{I}_{u,v}
    =
    \operatorname{TopK}_{B_{\rm corr}}
    \left(
    \left\{
    s_{\rm corr}
    (\mathbf{x}_{j};\mathbf{x}_{\rm t},\mathbf{x}_{u,v}(z))
    \right\}_{j}
    \right),
    \label{eq:ray_corridor}
\end{equation}
where $B_{\rm corr}$ is the local memory budget. The score combines geometric
proximity to the transmitter--receiver segment and endpoint neighborhoods; the
default budget is reported in Appendix~\ref{app:model_architecture}.

For each selected token, the decoder forms an augmented local feature
\begin{equation}
    \boldsymbol{\psi}^{u,v}_{j}
    =
    \psi
    \left(
    \mathbf{X}^{\rm loc}_{j},
    \mathbf{x}_{j},
    \mathbf{x}_{\rm t},
    \mathbf{x}_{u,v}(z)
    \right),
    \label{eq:psi_token}
\end{equation}
where $\psi(\cdot)$ appends normalized geometric relation features to the
selected sparse scene token. The receiver-specific memory is then
\begin{equation}
    \mathbf{M}^{\rm rm}_{u,v}
    =
    \operatorname{Concat}
    \left(
    \left\{
    \boldsymbol{\psi}^{u,v}_{j}
    \mid
    j\in\mathcal{I}_{u,v}
    \right\},
    \mathbf{M}_{\rm tx}
    \right).
    \label{eq:rm_memory}
\end{equation}

Each receiver query cross-attends to its receiver-specific memory
$\mathbf{M}^{\rm rm}_{u,v}$. The resulting receiver-plane tokens then exchange
spatial context through self-attention, are reshaped into a two-dimensional
feature map, and are processed by a convolutional head to produce the dense
path-gain map $\widehat{G}_{\theta}(\mathbf{x}_{\rm t},z)$, as shown in
Fig.~\ref{fig:overview}(b).

\subsection{Multipath CIR Set Decoder}
\label{sec:cir_decoder}

Multipath CIR prediction is a path-level set prediction problem. For a
transmitter--receiver pair $(\mathbf{x}_{\rm t},\mathbf{x}_{\rm r})$, the
decoder predicts a variable-cardinality set of delay--power taps. Since
physical paths have no canonical ordering, imposing a fixed sequence order
would introduce an artificial supervision constraint. We therefore use a
DETR-style set decoder with $K$ learnable candidate
tap queries~\cite{detr}.

Let $\mathbf{u}_{k}\in\mathbb{R}^{C}$ denote the $k$-th learnable tap query,
for $k=1,\ldots,K$. The receiver location conditions the initial query as
\begin{equation}
    \mathbf{q}^{(0)}_k
    =
    \mathbf{u}_k
    +
    \operatorname{Proj}_{\rm rx}
    \left(
    \operatorname{PE}(\mathbf{x}_{\rm r})
    \right),
    \label{eq:cir_query_init}
\end{equation}
where $\operatorname{PE}(\cdot)$ denotes the positional encoding. The
receiver-conditioned query bank attends to $\mathbf{M}_{\rm tx}$ and is decoded
into candidate taps
\begin{equation}
    \widehat{\mathbf{y}}_k
    =
    \left(
    \widehat{\tau}_k,
    \widehat{\gamma}_k,
    \widehat{e}_k
    \right),
    \qquad k=1,\ldots,K,
    \label{eq:pred_tap}
\end{equation}
where $\widehat{\tau}_k$ is the predicted delay, $\widehat{\gamma}_k$ is the
predicted tap power in dB, and $\widehat{e}_k$ is an existence logit. Bounded
output heads restrict delay and power to dataset-defined ranges, while the
existence logit allows unused query slots to represent no-path predictions.
The decoder therefore models the CIR as an unordered sparse path set rather
than as a fixed ordered vector. This set-decoding structure is summarized in
Fig.~\ref{fig:overview}(c).

\subsection{Training Objective and Optimization}
\label{sec:training}

The training objective combines dense radiomap supervision with
permutation-invariant CIR set supervision:
\begin{equation}
    \mathcal{L}
    =
    \mathcal{L}_{\rm rm}
    +
    \lambda_{\rm cir}
    \mathcal{L}_{\rm cir}.
    \label{eq:joint_loss}
\end{equation}

For radiomap prediction, let
$\Omega=\{(u,v):M_{u,v}=1\}$ denote the valid receiver cells. The radiomap loss
is
\begin{align}
    \mathcal{L}_{\rm rm}
    &=
    \frac{1}{|\Omega|}
    \sum_{(u,v)\in\Omega}
    \operatorname{Huber}_{\beta}
    \left(
    \widehat{G}_{u,v}-G^{*}_{u,v}
    \right)
    \nonumber\\
    &\quad+
    \lambda_{\rm grad}
    \left\|
    \nabla_{\Omega}\widehat{G}
    -
    \nabla_{\Omega}G^{*}
    \right\|_{1},
    \label{eq:rm_loss}
\end{align}
where $G^{*}$ is the ground-truth path-gain map, $M_{u,v}$ is the valid-cell
mask, and $\beta$ is the Huber threshold. The gradient term encourages spatial
consistency over valid regions.

For CIR prediction, the predicted tap set
$\{\widehat{\mathbf{y}}_k\}_{k=1}^{K}$ is matched to the ground-truth set
$\mathcal{Y}^{*}=\{(\tau_m,\gamma_m)\}_{m=1}^{M}$ using Hungarian matching.
The matching cost between ground-truth tap $m$ and candidate $k$ is
\begin{equation}
    C_{m,k}
    =
    a_{\tau}
    \left|
    \widehat{\tau}_{k}-\tau_m
    \right|
    +
    a_{\gamma}
    \left|
    \widehat{\gamma}_{k}-\gamma_m
    \right|
    -
    a_e
    \log
    \sigma(\widehat{e}_k),
    \label{eq:matching_cost}
\end{equation}
where $a_{\tau}$, $a_{\gamma}$, and $a_e$ are matching weights and
$\sigma(\cdot)$ is the sigmoid function. Let $\pi^{\star}$ denote the
minimum-cost assignment from ground-truth taps to prediction slots. The CIR
loss is
\begin{align}
    \mathcal{L}_{\rm cir}
    &=
    \sum_{m=1}^{M}
    \left(
    a_{\tau}
    \left|
    \widehat{\tau}_{\pi^{\star}(m)}-\tau_m
    \right|
    +
    a_{\gamma}
    \left|
    \widehat{\gamma}_{\pi^{\star}(m)}-\gamma_m
    \right|
    \right)
    \nonumber \\
    &\quad+
    \lambda_{\rm exist}
    \operatorname{BCE}
    \left(
    \widehat{\mathbf{e}},
    \mathbf{e}^{*}
    \right).
    \label{eq:cir_loss}
\end{align}
The existence target $e^{*}_{k}$ is one for matched slots and zero otherwise.
Matched slots are supervised by delay and peak-power regression, while
unmatched slots are supervised as no-path predictions through the existence
term. Here, $\operatorname{BCE}(\cdot)$ denotes the binary cross-entropy loss.
The numerical weights and Huber thresholds used in the default run are
reported in Appendix~\ref{app:model_architecture}.

Direct joint optimization from random initialization can be unstable because
the two views have different output dimensions, loss scales, and convergence
rates. We therefore use a warm-started alternating strategy consisting of
task-specific initialization, a short joint alignment phase, and alternating
task-focused updates. During each task-focused phase, the shared encoder and
the active decoder are updated while the inactive decoder is held fixed.
Detailed schedules and hyperparameters are reported in Section~\ref{sec:results}.

The encoder is evaluated once for each scene--transmitter pair. The resulting
memory can be reused for all receiver-plane radiomap queries and for multiple
receiver-point CIR queries. If $N_{\rm occ}$ is the number of occupied voxels,
the ray-corridor gather bounds fine-scale attention by $B_{\rm corr}$ rather
than $N_{\rm occ}$. The CIR decoder uses a fixed number of candidate taps $K$,
so its cost is independent of the unknown number of physical paths.
Consequently, the framework amortizes scene-and-transmitter encoding across
heterogeneous wireless queries while allowing each wireless view to retain its
own query structure, memory access pattern, and loss.

\section{Co-Registered Scene--Wireless Dataset Generation}
\label{sec:dataset_generation}

Training and evaluating WiSER requires a data interface in which one physical
scene representation supports multiple wireless supervision views. A central
contribution of this work is therefore a co-registered indoor scene--wireless
dataset pipeline for geometry-conditioned wireless representation learning.
Existing wireless-learning datasets are typically organized around a single
output representation, such as radiomaps, channel knowledge maps, or link-level
channel responses, and often differ in scene format, coordinate convention,
propagation model, and query interface. To the best of our knowledge, existing
indoor datasets do not provide the specific combination required here: sparse
3D scene inputs, dense radiomap labels, and unordered multipath CIR tap sets aligned under one coordinate frame and ray-tracing
configuration.

\begin{figure*}[t]
  \centering
  \includegraphics[width=0.99\textwidth]{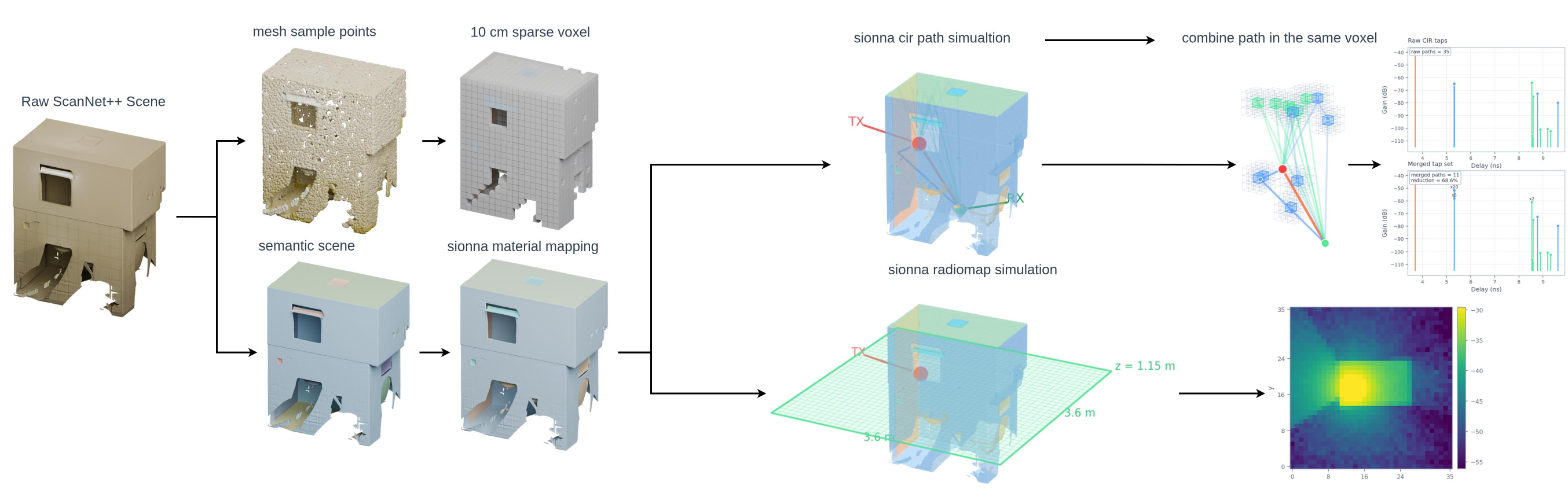}
  \caption{Co-registered dataset-generation pipeline. A ScanNet++ scene is
  converted into a sparse voxel scene for learning and a Sionna-compatible
  radio scene for simulation. The same geometry and coordinate frame produce
  aligned dense radiomap labels and path-level CIR labels; raw multipath
  responses are consolidated into compact unordered tap sets for set
  prediction.}
  \label{fig:dataset_pipeline}
\end{figure*}

The proposed pipeline addresses this gap by generating complementary wireless
views from the same physical scene. As shown in
Fig.~\ref{fig:dataset_pipeline}, a ScanNet++ indoor scene~\cite{scannetpp} is
converted into two aligned products: a sparse voxel representation used by
WiSER and a Sionna Ray Tracing scene used for label generation~\cite{sionna_rt}.
Dense radiomap labels, path-level CIR labels, transmitter and receiver
coordinates, valid masks, and metadata are produced from the same radio scene.
This alignment is essential because WiSER is designed to learn a shared scene
representation rather than independent single-task predictors. The following
subsections describe the co-registration principle, sparse-scene and radio-scene
construction, ray-tracing label generation, and multipath consolidation.

\subsection{Co-Registration Principle}
\label{sec:dataset_overview}

For the $s$-th indoor scene, let $\mathcal{S}_s$ denote the physical scene and
let $\mathcal{V}_s$ denote its sparse voxel representation. Candidate
transmitter locations are denoted by
$\mathcal{T}_s=\{\mathbf{x}^{(s)}_{{\rm t},m}\}_{m=1}^{N_{\rm tx}^{(s)}}$.
For a receiver plane at height $z$, the receiver lattice is
\begin{equation}
    \mathcal{R}_s(z)
    =
    \left\{
    \mathbf{x}^{(s)}_{u,v}(z)
    \;\middle|\;
    1\leq u\leq N_y^{(s)}(z),\;
    1\leq v\leq N_x^{(s)}(z)
    \right\},
    \label{eq:dataset_rx_lattice}
\end{equation}
where $\mathbf{x}^{(s)}_{u,v}(z)$ is the receiver coordinate associated with
grid cell $(u,v)$, and $N_y^{(s)}(z)$ and $N_x^{(s)}(z)$ are the row and
column counts of the receiver lattice for scene $s$ and plane $z$.

The co-registered interface is designed around three requirements:
geometric alignment between the sparse learning input and the radio scene,
wireless-label alignment across radiomap and CIR views, and target consistency
between continuous ray-tracing paths and the voxel-resolution scene
representation. Accordingly, all scene inputs, transmitter and receiver
coordinates, radiomap cells, ray-traced paths, masks, and metadata are defined
under a common scene coordinate system and propagation configuration.

A radiomap example consists of the shared scene input $\mathcal{V}_s$, a
transmitter location $\mathbf{x}_{\rm t}$, receiver-plane height $z$, dense
path-gain label $G_s(\mathbf{x}_{\rm t},z)$, and valid-cell mask
$M_s(\mathbf{x}_{\rm t},z)$. A CIR example consists of the same scene input, a
transmitter--receiver pair $(\mathbf{x}_{\rm t},\mathbf{x}_{\rm r})$, and an
unordered tap set $\mathcal{Y}_s(\mathbf{x}_{\rm t},\mathbf{x}_{\rm r})$.
Thus, both label types are linked to the same physical scene representation
and can supervise a single geometry-conditioned model.

All wireless labels in this work are generated by Sionna Ray Tracing rather
than over-the-air measurements. The dataset therefore provides controlled
simulator supervision for studying scene-conditioned representation learning,
rather than measurement-calibrated real-world channel ground truth.

\subsection{Sparse Scene and Radio-Scene Construction}
\label{sec:dataset_voxelization}

Each ScanNet++ scene provides high-fidelity geometry, RGB
texture, and semantic annotations, which are used to construct both sparse 3D
learning inputs and Sionna-compatible radio scenes with semantic-to-material assignments. Although
the sparse voxel interface is modality-agnostic, the instantiation in this
paper uses ScanNet++ geometry and semantics to construct the voxel features and
simulation scene.

The learning input is constructed by sampling the scene surface and voxelizing
the resulting 3D observations at voxel edge length $c_v$. Only occupied surface
or object voxels are retained, producing
\begin{equation}
    \mathcal{V}_s
    =
    \left\{
    (\mathbf{p}_i,\mathbf{f}_i)
    \right\}_{i=1}^{N_s},
    \label{eq:dataset_voxel_set}
\end{equation}
consistent with the sparse voxel representation in
Section~\ref{sec:background}. The feature vector $\mathbf{f}_i$ includes
occupancy, averaged color, semantic category, and material-related attributes.
This sparse representation is the common scene input shared by the radiomap
and CIR views; it preserves propagation-relevant surfaces and objects while
avoiding dense computation over empty indoor volume.

The same scene geometry is converted into a Sionna-compatible radio scene for
label generation. Semantic categories are mapped to a finite set of simulator
material classes so that walls, floors, ceilings, windows, furniture, and other
objects induce different simulated propagation responses. This
semantic-to-material mapping is a controlled simulation convention rather than
measured radio-frequency material identification. Since all labels and all
evaluated models use the same mapping, it provides a consistent supervision
model for comparing geometry-conditioned predictors. The sparse voxel scene and
the radio scene are therefore co-registered by construction.

\subsection{Ray-Tracing Labels for Radiomap and CIR Views}
\label{sec:dataset_sionna}

Radiomap and CIR labels are generated from the same Sionna Ray Tracing scene.
For each experimental instantiation, the carrier frequency, antenna
configuration, maximum interaction depth, enabled propagation mechanisms, and
material convention are fixed and stored with the dataset metadata.

\emph{Radiomap labels.}
For transmitter $\mathbf{x}_{\rm t}\in\mathcal{T}_s$ and receiver plane $z$,
the coverage solver produces a dense path-gain field
$G_s(\mathbf{x}_{\rm t},z)\in\mathbb{R}^{N_y^{(s)}(z)\times N_x^{(s)}(z)}$ over the
lattice $\mathcal{R}_s(z)$. For each valid receiver cell,
\begin{equation}
    G_s(\mathbf{x}_{\rm t},z)[u,v]
    =
    10\log_{10}
    \left(
    \sum_{\ell\in
    \mathcal{L}_s(\mathbf{x}_{\rm t},\mathbf{x}^{(s)}_{u,v}(z))}
    |\alpha_{\ell}|^2
    \right),
    \label{eq:dataset_radiomap_label}
\end{equation}
where $\mathcal{L}_s(\mathbf{x}_{\rm t},\mathbf{x}^{(s)}_{u,v}(z))$ is the
set of ray-traced paths reaching the receiver cell and $\alpha_{\ell}$ is the
complex path coefficient. This label is a path-gain quantity in dB under the
normalization convention used for the path coefficients; transmit-power
dependent received power can be obtained by adding the transmit power under
the same convention. A valid-cell mask $M_s(\mathbf{x}_{\rm t},z)$ is stored
with each radiomap and propagated to the training and evaluation interfaces.

\emph{CIR labels.}
For a transmitter--receiver pair
$(\mathbf{x}_{\rm t},\mathbf{x}_{\rm r})$, the path solver returns a raw
multipath response. Each path record contains delay, complex path coefficient,
departure and arrival angular descriptors, ordered interaction labels, and
ordered bounce-point geometry. The supervised CIR target in this paper uses
delay and power; the remaining metadata is retained for reproducibility and for
future extensions involving angular or interaction-level wireless views. The
delay reference is fixed across all generated samples and is applied
consistently to both training labels and evaluation metrics.

Because both label types are generated from the same radio scene, they provide
aligned field-level and path-level supervision for the same scene-conditioned
channel.

\subsection{Multipath Consolidation and Dataset Interface}
\label{sec:dataset_path_merge}

Raw ray-tracing outputs may contain multiple paths that differ only at a
spatial scale finer than the sparse voxel input. Directly supervising the CIR
set decoder with all such paths would require the model to distinguish
propagation events that are not resolved by its scene representation. We
therefore consolidate paths that share the same interaction depth,
interaction-type sequence, and quantized bounce-voxel sequence. This
consolidation is used only to construct supervised tap targets at the
resolution of the sparse voxel input; it is not intended to replace coherent
field summation in the simulator. The original complex ray-tracing outputs are
retained in the raw-path metadata, while the noncoherent merge is used only for
supervised target construction.

For raw path $\ell$, let
$\mathcal{B}_{\ell}
=
(\mathbf{b}_{\ell}^{(1)},\ldots,\mathbf{b}_{\ell}^{(D_{\ell})})$
denote its ordered bounce-point sequence, where $D_{\ell}$ is the interaction
depth. With scene origin $\mathbf{x}_0$ and voxel edge length $c_v$, each
bounce point is quantized as
\begin{equation}
    \widetilde{\mathbf{b}}_{\ell}^{(d)}
    =
    \left\lfloor
    \frac{\mathbf{b}_{\ell}^{(d)}-\mathbf{x}_0}{c_v}
    \right\rfloor .
    \label{eq:bounce_quantization}
\end{equation}
The path identity key is
\begin{equation}
    \mathcal{K}(\ell)
    =
    \left(
    D_{\ell},
    \boldsymbol{\kappa}_{\ell},
    \widetilde{\mathbf{b}}_{\ell}^{(1)},\ldots,
    \widetilde{\mathbf{b}}_{\ell}^{(D_{\ell})}
    \right),
    \label{eq:path_key}
\end{equation}
where $\boldsymbol{\kappa}_{\ell}$ is the ordered interaction-type sequence.
Paths are grouped only when they share the same key.

For an equivalence class $\mathcal{G}$ of grouped paths, one representative tap
is emitted. Its delay is the power-weighted average, and its power is the
accumulated noncoherent path power:
\begin{align}
    P_{\mathcal{G}}
    &=
    \sum_{\ell\in\mathcal{G}}
    |\alpha_{\ell}|^2,
    \label{eq:group_power} \\
    \tau_{\mathcal{G}}
    &=
    \frac{1}{P_{\mathcal{G}}}
    \sum_{\ell\in\mathcal{G}}
    |\alpha_{\ell}|^2\tau_{\ell},
    \label{eq:group_delay} \\
    \gamma_{\mathcal{G}}
    &=
    10\log_{10}
    P_{\mathcal{G}} .
    \label{eq:group_power_db}
\end{align}
The set of all representative taps forms
$\mathcal{Y}_s(\mathbf{x}_{\rm t},\mathbf{x}_{\rm r})$. This construction
preserves total noncoherent power within each voxel-bin equivalence class while
reducing redundant path multiplicity.
Each generated example is accompanied by metadata specifying the scene
identifier, coordinate transform, voxelization parameters, semantic-to-material
mapping, ray-tracing configuration, transmitter and receiver coordinates,
masks, and tap-construction parameters. This manifest-based design makes the
generated labels auditable and enables reproducible evaluation of whether a
single sparse 3D scene representation can support both dense field-level and
sparse path-level wireless prediction.

\section{Experimental Evaluation}
\label{sec:results}

We evaluate WiSER to answer four questions. First, can a single multi-scene
wireless scene encoder predict dense radiomap fields competitively against
scene-specific neural field baselines? Second, can the same scene representation
support path-level multipath CIR prediction under
delay--power supervision with Hungarian matching? Third, which architectural
components are responsible for the observed performance? Fourth, is
warm-started alternating optimization necessary for training a shared model
under heterogeneous dense-field and sparse-set supervision?

\subsection{Experimental Setup}
\label{sec:experimental_protocol}

The generic dataset construction is described in
Section~\ref{sec:dataset_generation}. This subsection specifies the concrete
evaluation setting, model instantiation, baselines, optimization settings, and
metrics used in the experiments.

\textbf{Dataset and evaluation.}
WiSER is trained on the 100-scene co-registered ScanNet++/Sionna dataset
described in Section~\ref{sec:dataset_generation}. The main radiomap and CIR
comparisons use ten evaluated scenes. For radiomap prediction, the evaluated
cases are formed by held-out transmitter locations and receiver-plane heights.
For CIR prediction, each evaluated scene uses held-out transmitter--receiver
triples. All methods use the same radiomap masks. CIR taps are compared after
Hungarian matching between the predicted and ground-truth tap sets.

\textbf{Model instantiation.}
Unless otherwise specified, all experiments use the default WiSER
configuration: a transmitter-conditioned sparse scene encoder, a ray-corridor
radiomap decoder, and a DETR-style CIR set decoder. The default model uses a
fixed CIR query budget of $K=8$. Layer counts, hidden widths, ray-corridor
budgets, output bounds, and loss settings are summarized in
Appendix~\ref{app:model_architecture}.

\textbf{Baselines.}
For radiomap prediction, we compare against NeRF2~\cite{nerf2} and
radio-frequency 3D Gaussian splatting (RF-3DGS)~\cite{rf3dgs}. These baselines
are trained separately for each evaluated scene, whereas WiSER is trained once
on the multi-scene training set and reused across all evaluated scenes. This
comparison therefore favors the baselines in terms of scene-specific
adaptation. For CIR prediction, existing public baselines are not directly
aligned with the combination of sparse 3D scene input, transmitter--receiver
conditioning, variable-cardinality tap prediction, and set matching considered
here. We therefore use three reference baselines of increasing capacity: Ridge
regression using geometric link features, an MLP using the same geometry-only
input, and a 3D convolutional neural network (CNN) using
scene voxels together with transmitter--receiver coordinates.

\textbf{Optimization.}
Internal comparisons include single-task radiomap training, single-task CIR
training, direct joint training from random initialization, and the proposed
warm-started alternating schedule. WiSER is initialized from single-task
radiomap and CIR checkpoints, followed by a short joint warm-up and alternating
task-focused phases. During each task-focused phase, the shared encoder and the
active decoder are updated while the inactive decoder is held fixed. Training
uses bfloat16 distributed data parallelism on eight NVIDIA A100 GPUs.

\textbf{Metrics.}
Radiomap prediction is evaluated over valid receiver cells using mean absolute
error (MAE), root mean squared error (RMSE), and peak signal-to-noise ratio
(PSNR). For a radiomap dynamic range $\Delta_G$, PSNR is computed as
$20\log_{10}(\Delta_G/\mathrm{RMSE})$; in our reported tables,
$\Delta_G=120$~dB. CIR prediction is evaluated after Hungarian matching between
predicted and ground-truth tap sets. We report matched peak-power MAE, matched
delay MAE, and path-count accuracy. Path-count accuracy is computed by
thresholding predicted existence probabilities with a fixed threshold
$\eta_{\rm exist}$ and comparing the predicted number of active taps with the
ground-truth tap count. Tables that explicitly report training-split results
use the same metric definitions on the corresponding training samples to
isolate fitting and optimization behavior.

\subsection{Radiomap Prediction}
\label{sec:results_rm}

Table~\ref{tab:table1_rm_compare} reports radiomap prediction results on ten
evaluated scenes, using held-out transmitter locations and receiver-plane
heights from the evaluation split. WiSER obtains the lowest radiomap error,
with 3.834~dB MAE
and 5.500~dB RMSE. Compared with the stronger RF-3DGS baseline, WiSER reduces
MAE by 0.75~dB and improves PSNR by 1.16~dB. NeRF2 has larger error on this
set, indicating that the radiomap task benefits from explicit 3D scene
conditioning.

\begin{table}[t]
\caption{Radiomap comparison.}
\label{tab:table1_rm_compare}
\centering
\begin{tabular}{lccc}
\toprule
Method & MAE (dB) $\downarrow$ & RMSE (dB) $\downarrow$ & PSNR (dB) $\uparrow$ \\
\midrule
NeRF2 & 8.238 & 10.056 & 21.54 \\
RF-3DGS & 4.585 & 6.281 & 25.62 \\
\textbf{WiSER} & \textbf{3.834} & \textbf{5.500} & \textbf{26.78} \\
\bottomrule
\end{tabular}
\end{table}

\begin{figure*}[t]
\centering
\includegraphics[width=0.98\textwidth]{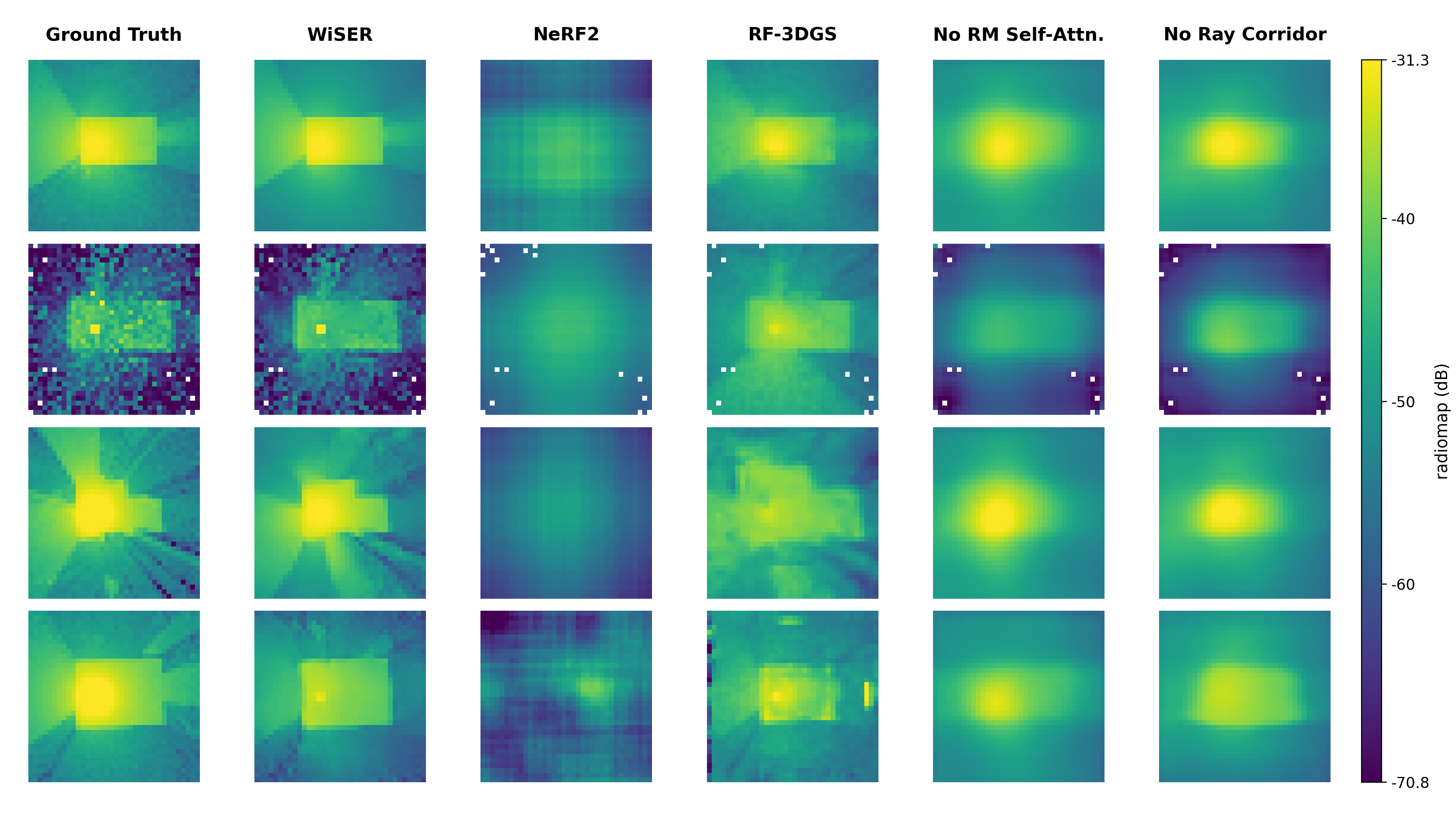}
\caption{Qualitative radiomap comparison on four representative scene--TX
cases. Columns show ground truth,
WiSER, NeRF2, RF-3DGS, and two radiomap-head ablations that remove receiver
query self-attention or ray-corridor voxel gathering. Rows include both
seen-height and held-out-height cases. Across these examples, WiSER better
preserves dominant attenuation lobes and geometry-induced shadow boundaries,
whereas the ablated variants tend to produce smoother maps under the same dB
color scale.}
\label{fig:qual_rm}
\end{figure*}

The comparison is conservative with respect to WiSER because NeRF2 and RF-3DGS
are optimized separately for each evaluated scene, while WiSER is trained once
and reused across all test cases. The lower MAE and RMSE suggest that WiSER
learns a reusable sparse scene-conditioned representation for indoor coverage
prediction rather than only fitting one scene at a time. The qualitative
comparison in Fig.~\ref{fig:qual_rm} is consistent with the quantitative
metrics. WiSER better preserves dominant high-power regions and
geometry-induced attenuation boundaries, whereas the neural field baselines tend
to produce smoother fields or localized artifacts under the shared dB scale.

\subsection{Multipath CIR Prediction}
\label{sec:results_cir}

For CIR prediction, the reference baselines are trained per evaluated scene and
tested on held-out transmitter--receiver triples from the same scene. Ridge
regression and the MLP use the same 8-dimensional geometric link feature vector,
consisting of transmitter coordinates, receiver coordinates, link distance, and
height difference. Ridge regression predicts the fixed-budget delay, peak-power,
and existence outputs using independent linear regressors, while the MLP uses a
three-layer nonlinear predictor on the same geometry-only input. The 3D CNN
baseline additionally processes the voxelized scene before concatenating scene
features with transmitter--receiver features. All three baselines use the same
fixed-budget output convention and are evaluated with the same matching-based
metrics.

Each reference baseline is trained on 2500 triples with transmitters
0-9 per scene and evaluated on 2500 held-out triples with transmitters
17-26 from the same scene. Table~\ref{tab:table2_csi_compare} reports
the mean over ten evaluated scenes. WiSER is trained once on 100 scenes
and evaluated by direct inference on the same held-out test triples.

\begin{table}[t]
\caption{Multi-path CIR comparison.}
\label{tab:table2_csi_compare}
\centering
\setlength{\tabcolsep}{3pt}
\footnotesize
\resizebox{\columnwidth}{!}{%
\begin{tabular}{lccc}
\toprule
Method & Peak-Power MAE (dB) & Delay MAE (ns) & Count Accuracy \\
       & $\downarrow$  & $\downarrow$   & $\uparrow$ \\
\midrule
Ridge regression & 163.69 & 4.76 & 0.382 \\
MLP & 21.16 & 2.84 & 0.410 \\
3D CNN & 11.50 & 1.50 & 0.407 \\
\midrule
\textbf{WiSER} & \textbf{5.89} & \textbf{0.61} & \textbf{0.477} \\
\bottomrule
\end{tabular}%
}
\end{table}

WiSER achieves a matched peak-power MAE of 5.89~dB and a matched delay MAE of
0.61~ns. Compared with the strongest reference baseline, the 3D CNN, WiSER
reduces peak-power error from 11.50~dB to 5.89~dB and delay error from
1.50~ns to 0.61~ns. It also achieves the highest path-count accuracy among the
evaluated methods.

The baseline progression clarifies the roles of scene geometry and structured
set prediction. Moving from the geometry-only MLP to the 3D CNN reduces
peak-power and delay errors, showing that scene information is useful for
path-level channel prediction. However, the 3D CNN still predicts a fixed
vector and does not explicitly align predictions with unordered propagation
taps. WiSER further improves matched delay and peak-power accuracy by using a
set decoder with Hungarian matching. Count accuracy improves more modestly,
which indicates that exact tap cardinality remains harder than matched delay
and power regression.

\begin{figure*}[t]
\centering
\includegraphics[width=0.99\textwidth]{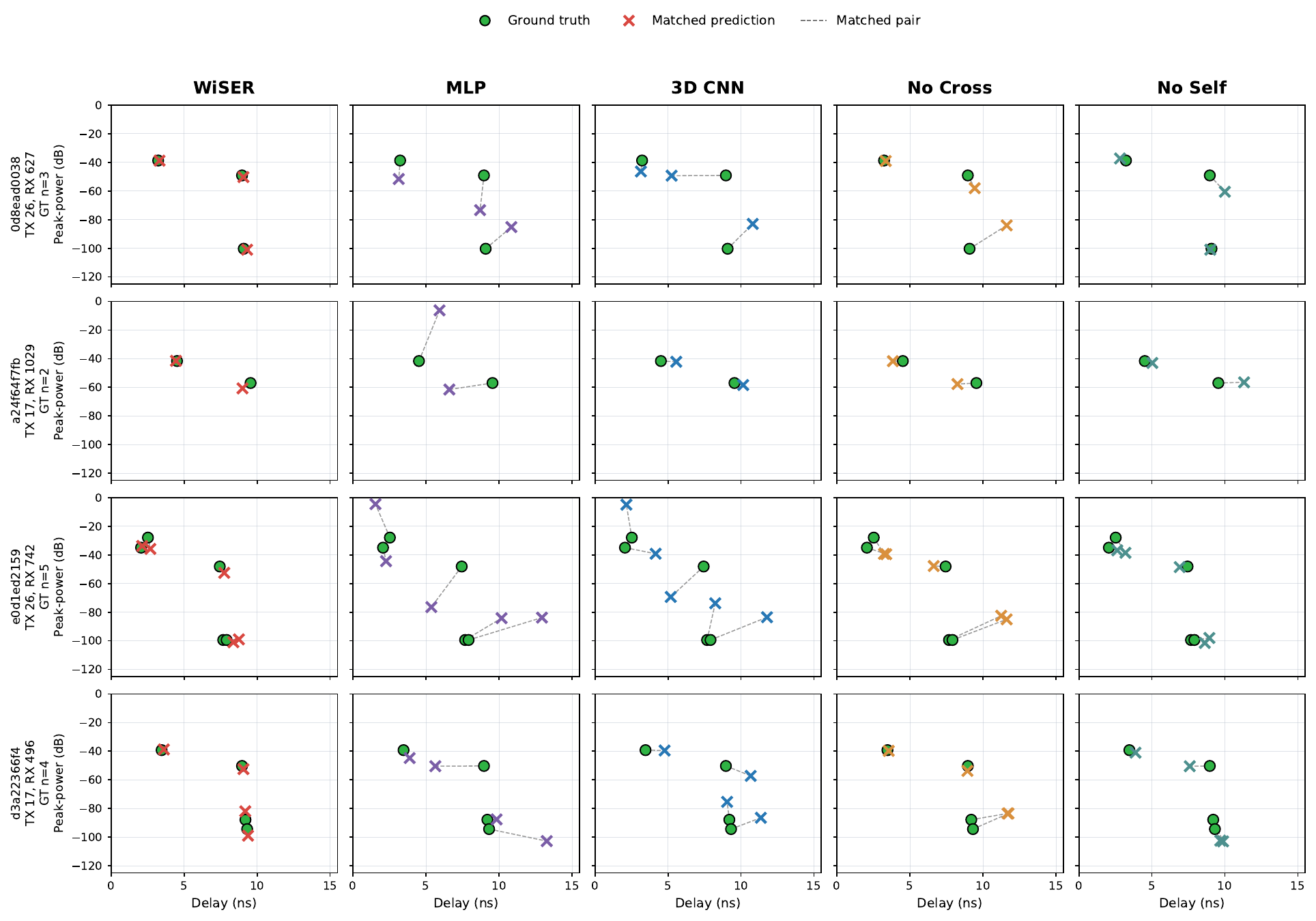}
\caption{Qualitative multipath CIR prediction examples with reference
baselines and attention ablations. Green circles denote ground-truth path taps,
colored crosses denote matched predicted taps, and dashed lines indicate
Hungarian matching pairs in the delay/peak-power plane. WiSER is compared with
the MLP and 3D CNN baselines, followed by CIR decoder variants without scene
cross-attention or query self-attention. Aggregate baseline and ablation metrics
are reported in Tables~\ref{tab:table2_csi_compare} and
\ref{tab:table3_arch_ablation}.}
\label{fig:cir_qualitative}
\end{figure*}

\subsection{Component Ablation}
\label{sec:results_ablation_arch}

The ablation study is intended to isolate component behavior under matched
training conditions. We use a reduced-scale single-branch setting with a
50-scene subset of the full training data. Radiomap-branch variants are trained
with the CIR loss disabled, and CIR-branch variants are trained with the
radiomap loss disabled. For each branch, the corresponding full-architecture
variant is evaluated as the reference. Table~\ref{tab:table3_arch_ablation}
reports training-split inference metrics on the corresponding samples
(4050 radiomap samples and 250000 CIR triples). Tables~\ref{tab:rm_ablation_within_scene_test}
and~\ref{tab:cir_ablation_within_scene_test} additionally report held-out
diagnostics within the same scene set: unseen receiver planes for radiomap
prediction and unseen TX-RX samples for CIR prediction.






\begin{table}[t]
\caption{Model component ablation on the training split.}
\label{tab:table3_arch_ablation}
\centering

\setlength{\tabcolsep}{4pt}
\renewcommand{\arraystretch}{1.08}

\footnotesize

\begin{tabular}{lccc}
\toprule

\multicolumn{4}{l}{\textit{Radiomap branch}} \\
\midrule

Variant 
& MAE $\downarrow$
& RMSE $\downarrow$
& \\

& (dB)
& (dB)
& \\

\midrule

Full model 
& \textbf{2.127}
& \textbf{3.539}
& \\

w/o radiomap self-attn.
& 2.971
& 4.597
& \\

w/o ray-corridor gather
& 2.430
& 3.953
& \\

\midrule

\multicolumn{4}{l}{\textit{Multi-path CIR branch}} \\
\midrule

Variant
& Peak $\downarrow$
& Delay $\downarrow$
& Count $\uparrow$ \\

& (dB)
& (ns)
& \\

\midrule

Full model
& \textbf{3.271}
& \textbf{0.250}
& \textbf{0.592} \\

w/o CIR cross-attn.
& 9.350
& 1.006
& 0.419 \\

w/o CIR self-attn.
& 6.861
& 1.089
& 0.434 \\

\bottomrule
\end{tabular}

\end{table}

\begin{table}[t]
\caption{Radiomap ablation on the within-scene held-out split.}
\label{tab:rm_ablation_within_scene_test}
\centering
\setlength{\tabcolsep}{5pt}
\footnotesize
\begin{tabular}{lcc}
\toprule
Variant & MAE (dB) $\downarrow$ & RMSE (dB) $\downarrow$ \\
\midrule
Full model & \textbf{3.779} & \textbf{5.666} \\
w/o radiomap self-attention & 4.467 & 6.215 \\
w/o ray-corridor gather & 3.899 & 5.785 \\
\bottomrule
\end{tabular}
\end{table}

\begin{table}[t]
\caption{CIR ablation on within-scene held-out TX-RX samples.}
\label{tab:cir_ablation_within_scene_test}
\centering
\setlength{\tabcolsep}{4pt}
\footnotesize
\begin{tabular}{lccc}
\toprule
Variant & Peak MAE & Delay MAE & Count Accuracy \\
        & (dB) $\downarrow$ & (ns) $\downarrow$ & $\uparrow$ \\
\midrule
Full model & \textbf{5.693} & 1.417 & \textbf{0.364} \\
w/o CIR cross-attention & 8.654 & \textbf{1.011} & 0.000 \\
w/o CIR self-attention & 8.011 & 1.380 & 0.069 \\
\bottomrule
\end{tabular}
\end{table}

On the radiomap branch, removing receiver-plane self-attention increases MAE
from 2.13~dB to 2.97~dB, while removing the ray-corridor gather increases MAE
to 2.43~dB on the training split. The within-scene held-out z-plane split
shows the same ordering: removing receiver-plane self-attention increases MAE to
4.47~dB, and removing ray-corridor gathering increases it to 3.90~dB. These
results indicate that both components help fit dense coverage fields and
retain useful behavior on unseen receiver planes within known scenes.
Receiver-plane self-attention supplies spatial context across nearby receiver
cells, while ray-corridor gathering provides geometry-constrained scene
evidence to each receiver query.

On the CIR branch, both attention operations in the DETR-style set decoder are
important. Removing scene cross-attention increases training peak-power MAE
from 3.27~dB to 9.35~dB and delay MAE from 0.25~ns to 1.01~ns, while reducing
count accuracy from 0.59 to 0.42. Removing query self-attention further
increases peak-power MAE to 6.86~dB and lowers count accuracy to 0.43. On the
within-scene held-out TX-RX samples, removing cross-attention increases
peak-power MAE to 8.65~dB and reduces count accuracy to zero, while removing
self-attention increases peak-power MAE to 8.01~dB and lowers count accuracy to
0.07. These results indicate that cross-attention is needed to retrieve
scene-conditioned path evidence and self-attention helps the candidate taps
coordinate as a set.

\subsection{Training Strategy}
\label{sec:results_training}

Finally, we compare training strategies using training-split fitting metrics.
This analysis is intended to isolate optimization behavior, since the goal is
to determine whether the shared encoder can be trained effectively under two
heterogeneous supervision signals. These results should therefore be
interpreted as optimization diagnostics rather than held-out generalization
metrics.


\begin{table*}[t]
\caption{Training-scheme comparison.}
\label{tab:table5_training}
\centering
\setlength{\tabcolsep}{3pt}
\footnotesize

\begin{tabular}{lcccc}
\toprule
 & Radiomap train MAE & Radiomap train RMSE & CIR train peak MAE & Count Accuracy \\
Strategy & (dB) $\downarrow$ & (dB) $\downarrow$ & (dB) $\downarrow$ & $\uparrow$ \\
\midrule
Single-task radiomap & 0.44 & 0.78 & --- & --- \\
Single-task CIR & --- & --- & 7.49 & 0.50 \\
Joint (equal weights, from scratch) & 3.36 & 4.46 & 8.79 & 0.50 \\
\textbf{Pretrain + Alternating} & 0.82 & 1.34 & 6.14 & 0.43 \\
\bottomrule
\end{tabular}%
\end{table*}

Table~\ref{tab:table5_training} shows that direct joint training from scratch
underfits both views. On the radiomap side, joint-from-scratch training obtains
3.36~dB MAE, whereas warm-started alternating training reduces this error to
0.82~dB and remains reasonably close to the radiomap-only model. On the CIR
side, warm-started alternating training achieves the lowest peak-power MAE
among the compared training strategies, improving over both the single-task CIR
model and the joint-from-scratch model. The joint-from-scratch model retains
slightly higher path-count accuracy, but it underfits both tasks more strongly
in terms of radiomap error and CIR peak-power calibration.

These results support the use of warm-started alternating optimization for
WiSER. Dense radiomap supervision and sparse CIR set supervision
have different output dimensions, loss scales, and convergence behavior.
Initializing each decoder with task-specific training and then alternating
task-focused updates allows the shared encoder to support both views without
the severe underfitting observed under direct joint training. The remaining gap
to the radiomap-only model reflects the cost of sharing capacity with CIR
prediction, while the gain over joint-from-scratch and the improvement over the
CIR-only model indicate that the shared representation is useful rather than
merely a compromise.

Taken together, the results support the central claim of WiSER: a single
transmitter-conditioned sparse 3D scene representation can serve as a shared
interface for both dense coverage-level prediction and sparse path-level
multipath prediction. The radiomap experiments demonstrate reusable
field-level prediction across scenes, the CIR experiments show that the same
memory supports structured delay--power tap prediction, and the diagnostics
indicate that output-specific decoding and warm-started alternating
optimization are both important for the unified model.

\section{Conclusion}
\label{sec:conclusion}
This paper studied whether a single geometry-conditioned 3D scene
representation can support both coverage-level and path-level wireless
prediction. We introduced a co-registered dataset-generation pipeline
that aligns sparse voxel scene inputs with Sionna-derived radiomap and
CIR supervision, and we developed WiSER, a wireless scene encoder with
two task-specific decoders: a ray-corridor radiomap decoder for dense
receiver-plane prediction and a DETR-style CIR decoder for
variable-cardinality multipath tap prediction. The experimental
results show that WiSER outperforms scene-specific radiomap
baselines on the evaluated scenes, substantially improves matched
CIR delay and peak-power prediction over reference CIR baselines, and
benefits from task-specific decoder designs such as ray-corridor
feature gathering and set-based CIR decoding. The training-scheme
study further shows that warm-started alternating optimization is more
effective than direct joint training from scratch for this heterogeneous
multi-task objective. These results support the central conclusion of
the paper: a shared sparse 3D scene representation can serve as a
common interface for multiple wireless propagation queries when paired
with decoders matched to the structure of each output space.




\appendices
\section{WiSER Module Details}
\label{app:model_architecture}

This appendix records the concrete module dimensions used by WiSER. The sparse
3D attention and sparse downsampling operators follow the TRELLIS-2
implementation~\cite{trellis2}; the transmitter-conditioned scene interface,
ray-corridor radiomap head, and multipath CIR set head are WiSER-specific.

\noindent\textbf{Scene encoder.}
The input is a sparse occupied-voxel set at $c_v=0.10$~m. For each occupied
cell we concatenate the voxel center, point count, and RGB color into a 7-D
feature and zero-pad it to the backbone width $C=512$; the integer voxel
coordinate is kept separately for sparse attention. The TX-free scene pass has
$S=3$ stages. Each stage contains three sparse transformer blocks with 8 heads
(64 channels/head) and a feed-forward network (FFN) of width $4C=2048$,
followed by mean sparse downsampling with factor 2. Thus the effective voxel
sizes are $0.10,0.20,0.40,$ and $0.80$~m.

The TX coordinate is Fourier encoded into $3{\times}8{\times}2=48$ channels
and mapped by $48\!\rightarrow\!128\!\rightarrow\!128$. The backbone then maps
this TX code through $128\!\rightarrow\!512\!\rightarrow\!512$ and applies
three TX-conditioned sparse transformer blocks on the coarsest scene memory.
In each adaptive layer normalization zero (AdaLN-zero) block, a sigmoid linear
unit (SiLU)--linear modulation produces $6C=3072$ parameters, split into six
512-D vectors: shift, scale, and gate for self-attention, and shift, scale,
and gate for the FFN. The output is
$\mathbf{M}_{\rm tx}\in\mathbb{R}^{N_3\times512}$ per TX. The 10-cm memory
$\mathbf{X}^{(0)}\in\mathbb{R}^{N_0\times512}$ is retained for local
ray-corridor retrieval, while $\mathbf{M}_{\rm tx}$ supplies global
TX-conditioned context.

\noindent\textbf{Radiomap head.}
For each TX and queried height plane, the radiomap head uses one query for each
$36{\times}36$ receiver cell, i.e., 1296 queries. The query encoder forms a
10-D vector from RX, TX, RX--TX displacement, and distance; Fourier features
with 6 bands expand it to 130 channels, followed by
$130\!\rightarrow\!512\!\rightarrow\!512$ and layer normalization (LayerNorm).
Explicit RX and TX Fourier position embeddings are added, giving queries
$\mathbf{Q}\in\mathbb{R}^{1296\times512}$. For each query, ray-corridor gather
keeps the top $B_{\rm corr}=192$ fine voxels from $\mathbf{X}^{(0)}$ using
radius $\rho=0.20+0.08L$~m around the TX--RX segment and endpoint radius
$\eta=0.35+0.10L$~m. Each selected token receives an 8-D relation embedding
processed by $8\!\rightarrow\!512\!\rightarrow\!512$. The local 192 tokens are
concatenated with the global $\mathbf{M}_{\rm tx}$ tokens. The token decoder
has four cross-attention blocks and two receiver-plane self-attention blocks,
all at width 512 with 8 heads and $512\!\rightarrow\!2048\!\rightarrow\!512$
FFNs.

After attention, a linear projection maps
$[B,1296,512]\!\rightarrow\![B,1296,256]$, which is reshaped to
$[B,256,36,36]$. The convolutional refinement head first concatenates the
valid-cell mask and two normalized coordinate channels, producing
$[B,259,36,36]$, then applies a $3{\times}3$ convolution
$259\!\rightarrow\!256$. The stem contains four residual convolutional blocks;
each block uses group normalization (GroupNorm), a $3{\times}3$ convolution
$256\!\rightarrow\!512$, Gaussian error linear unit (GELU), GroupNorm, and a
$3{\times}3$ convolution $512\!\rightarrow\!256$. The output head uses four
36-by-36 residual blocks of the same $256\!\rightarrow\!512\!\rightarrow\!256$
form, a stride-2 $3{\times}3$ down-convolution $256\!\rightarrow\!512$ to
$18{\times}18$, four bottleneck residual blocks with $3{\times}3$ convolutions
$512\!\rightarrow\!1024\!\rightarrow\!512$, a $3{\times}3$ up-projection
$512\!\rightarrow\!256$ followed by bilinear upsampling to $36{\times}36$, a
fusion convolution $3{\times}3$ from 512 concatenated channels to 256, three
post-fusion residual refinement blocks
$256\!\rightarrow\!512\!\rightarrow\!256$, and a final $1{\times}1$
convolution $256\!\rightarrow\!1$ to produce the radiomap in dB.

\noindent\textbf{Multipath CIR head.}
For each TX--RX pair, the CIR head uses the same
$\mathbf{M}_{\rm tx}\in\mathbb{R}^{N_3\times512}$ memory. The RX coordinate is
Fourier encoded into $3{\times}8{\times}2=48$ channels and projected by
$48\!\rightarrow\!512\!\rightarrow\!512$. This RX embedding is added to
$K=8$ learned path queries, forming a query tensor $[Q,8,512]$ for $Q$ TX--RX
triples. The DETR-style decoder has six transformer decoder layers. Each layer
uses 8-head query self-attention, 8-head cross-attention to
$\mathbf{M}_{\rm tx}$, and an FFN $512\!\rightarrow\!2048\!\rightarrow\!512$;
all attention heads have dimension 64. The decoded tensor remains
$[Q,8,512]$. The final checkpoint uses deep projection heads for all three CIR
outputs. For each path slot, each head applies LayerNorm, a residual MLP
$512\!\rightarrow\!1024\!\rightarrow\!512$, another LayerNorm, and an output
MLP $512\!\rightarrow\!512\!\rightarrow\!1$. The existence head emits a raw
logit, the delay head uses a tanh-affine map to $[0,15]$~ns, and the
peak-power head uses a tanh-affine map to $[-115,-7.5]$~dB. Hungarian matching
aligns the unordered 8 slots with the supervised delay--power tap set during
training; at inference, the existence logits choose the active predicted taps.

\bibliographystyle{IEEEtran}
\bibliography{references}

\end{document}